\documentclass[journal=jacsat,manuscript=article]{achemso}

\usepackage[version=3]{mhchem} 
\usepackage[T1]{fontenc}       
\usepackage{bm}
\usepackage{soul}
\usepackage{amsmath,amssymb,graphicx}

\author{Avinash Rustagi}
\email{arustag@ncsu.edu}
\author{Alexander F. Kemper}
\email{akemper@ncsu.edu}
\affiliation[North Carolina State University]
{Department of Physics, North Carolina State University, Raleigh, NC, USA}

\title[An \textsf{achemso} demo]
  {Theoretical phase diagram for the room temperature Electron-Hole Liquid in photo-excited quasi-2D monolayer MoS$_2$}

\abbreviations{EHL,BGR,2D}

\begin{document}

%
%
%
%
%

\begin{abstract}
 Strong correlations between electrons and holes can drive the existence of an electron-hole liquid (EHL) state, typically at high carrier densities and low temperatures. The recent emergence of quasi-2D monolayer transition metal dichalcogenides (TMDCs) provide ideal systems to explore the EHL state since ineffective screening of the out of plane field lines in these quasi-2D systems allows for stronger charge carrier correlations in contrast to conventional 3D bulk semiconductors and enabling the existence of the EHL at high temperatures.  Here we construct the phase diagram for the photo-induced first-order phase transition from a plasma of electron-hole pairs to a correlated EHL state in suspended monolayer MoS2. We show that quasi-2D nature of monolayer TMDCs and the ineffective screening of the out of plane field lines allow for this phase transition to occur at and above room temperature, thereby opening avenues for studying many-body phenomena without the constraint of cryogenics.
\end{abstract}

\section{Introduction}
The monolayer transition-metal dichalcogenides, two-dimensional graphene-like lattices, are enabling the study of fundamental physics in regimes that are not typically accessible.  One aspect of this is the unusually high exciton binding energy due to the quantum confinement and lack of screening of the electron-hole pairs, and another is the possibility of photo-doping to comparatively high excitation densities.  An unusual feature of the latter is that at the highest densities, the mixture of excitons and electron-hole plasma can undergo a phase transition into a
strongly correlated \textit{electron-hole liquid} state \cite{keldysh1968proceedings,Keldysh1986}, with a critical temperature roughly set by the exciton binding energy. 
In the TMDCs, due to the high binding energy, the critical temperature should be similarly high, with a simple estimate placing it above room temperature.  This prediction was recently confirmed in a landmark experiment \cite{kenan}, where a monolayer of MoS$_2$ was photo-doped to extremely high carrier densities, resulting in photoluminescence spectra showing the signatures of an electron-hole liquid (EHL).

The transition in to an EHL in photo-doped systems is a striking example of a true non-equilibrium phase transition.
It occurs at high densities, when the mixture of electron-hole plasma and excitons undergoes a Mott transition triggered by runaway dissociation of excitons \cite{zimmermann1988many}. The properties of the mixture are typically set by temperature and interactions, leading to a complex interplay between thermal/Mott dissociation and correlations, resulting in a rich phase diagram for the non-equilibrium photo excited electron-hole system. In addition, a recent experiment has shown that high magnetic field alters the materials band structure and thus can tune the conditions of EHL formation
\cite{BStabilized_EHL2016}. This opens up new avenues for studying the many-body liquid phase under the effects of magnetic field, strain, etc. for potential technological applications.

The two ingredients to EHL formation are long-lived photo-excited carriers and the lower ground state energy of the state compared to the exciton/plasma mixture. 
In indirect-gap semiconductors such as silicon, germanium, gallium phosphide, the lower ground state energy of EHL was attributed to the multi-valley nature of the bands, and the different effective masses of electrons and holes \cite{TKLO1974,EHLGaP1977};  additional inclusion of coupling to phonons was necessary to explain EHL formation in direct-gap semiconductors \cite{EHLPolarSC1977,MoriyaGaAs1974}. Until the discovery in the TMDCs, these constraints had limited the critical temperature to 165K in diamond \cite{ShimanoDiamondEHL2014}. The above room temperature transition in the TMDCs raises the possibility of cryogen-free devices based on the EHL state, where the optical and electric properties are markedly different from the underlying semiconductor.

In this work, we determine the phase diagram of photo- induced non-equilibrium phase transition in the quasi-2D monolayer MoS$_2$ via a calculation of the many-body grand potential using the linked cluster expansion. Our results indicate the formation of EHL state at high densities with a maximum critical temperature of $\sim$514.9 K and a critical density of $\sim$3.8 $\times 10^{11}$ cm$^{-2}$. We find that the high critical temperatures are in part due to the nearly two-dimensional nature of the material, where the poor screening of field lines outside of the material plays a critical role.
 
 \begin{figure*}
 \centering
 \includegraphics[scale=0.35]{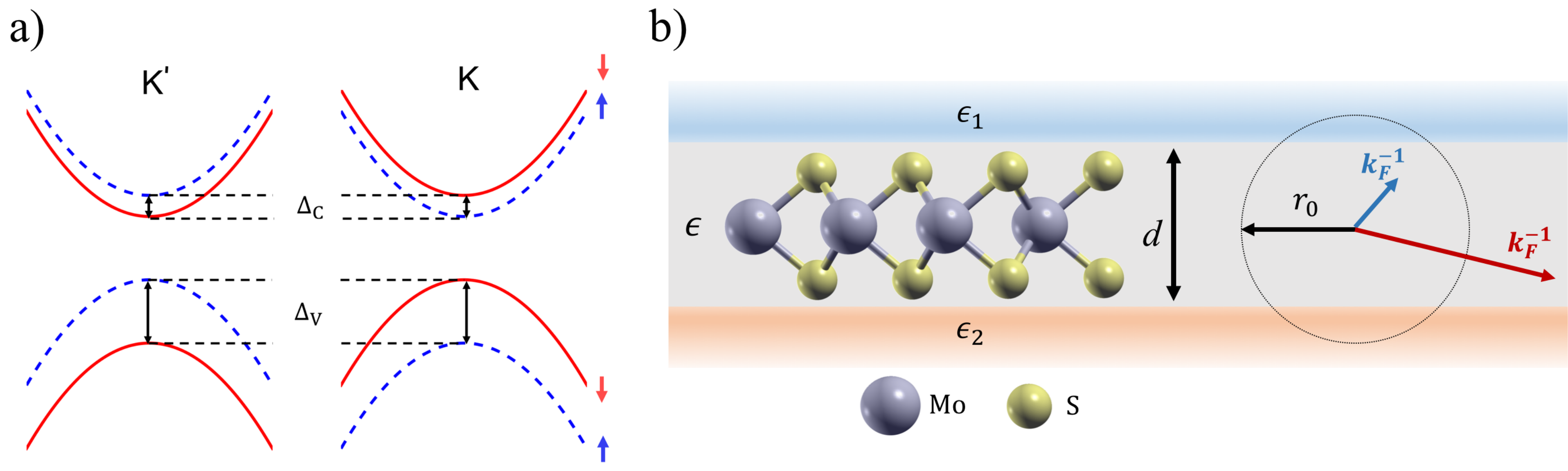}
 \caption{\label{bands} a) Monolayer MoS$_2$ band structure schematic displaying the Valence ($\Delta_\mathrm{V} \approx 148 $ meV) and Conduction ($\Delta_\mathrm{C} \approx 3 $ meV) band spin splitting due to spin-orbit coupling at the degenerate K and K' points in the Brillouin zone. b) Schematic of the quasi-2D monolayer between two substrates where charges interact via Keldysh potential characterized by length scale $r_0$. The interaction behaves as two-dimensional ($\sim 1/q$) for small inter-electron spacing (small $k_\mathrm{F}^{-1}$) and three-dimensional ($\sim 1/q^2$) for large inter-electron spacing (large $k_\mathrm{F}^{-1}$).}
 \end{figure*}
 
TMDCs provide a unique system to study many-body phenomena tuned by strain, magnetic field, etc. with the potential for a wide range of applications in optoelectronics \cite{Kis_Photodetector2013}, and valleytronics \cite{Heinz_Pseudospin2014,Heinz_ValleyPseudospin2017}. They are particularly versatile in tunability of band gap and direct/indirect gap nature of bands depending on the number of layers \cite{Shen_Indirect2Direct_2014}. The quasi-2D nature of the interactions, long-lived photo-excited carriers, and their particular band structure \textemdash a direct gap at the K and K' points \textemdash\ makes monolayer TMDCs an ideal class of candidates to study the EHL state. Strong spin-orbit coupling in these materials results in spin-splitting of the conduction and valence bands. The spin texture of the split bands at K and K' points are related by time reversal symmetry which is preserved for spin-orbit interaction. For MoS$_2$, the conduction band spin splitting is fairly small ($\sim$3 meV) and thus is usually ignored but the valence band spin splitting is significantly large ($\sim $ 148 meV). The valence band splitting of $\sim$ 148 meV implies that for a photo-excited density less than  $3.26 \times 10^{13}$ cm$^{-2}$, only one of the spin-split valence bands is occupied at each of the K and K' points as shown in Fig.~\ref{bands} which is what we include in our calculations.

\section{Ground State Energy}
The stability of EHL compared to excitons stems from its lower ground state energy which has contributions from kinetic ($E_\mathrm{k}$), exchange ($E_\mathrm{x}$), and correlation ($E_\mathrm{c}$) energies at T=0 K. These are given by the following expressions \cite{Hubbard1958,BrinkmanRice1973} :
\begin{equation}
\label{ET0}
\begin{split}
E_\mathrm{k} &= \sum_{i=e,h} \nu_{i}\sigma_i \sum_{\bm{k}<\bm{k}_{F,i}} \frac{\hbar^2 k^2}{2 m_{i}}\\
E_\mathrm{x} &= -\sum_{i=e,h}\frac{\nu_{i}\sigma_i}{2 L^2} \sum_{\bm{k},\bm{p}<\bm{k}_{F,i}} V_{\bm{k}- \bm{p} }\\
E_\mathrm{c} &= \sum_{\bm{q}} \int_0^\infty \frac{\hbar d\omega}{2 \pi} \left[ \tan^{-1} \left( \frac{-\Sigma_{\bm{q}}(\omega)}{1-A_{\bm{q}}(\omega)}\right) + \Sigma_{\bm{q}}(\omega) \right]
\end{split}
\end{equation}
where index $i=\{e,h\}$ identifies the electron and hole species, $\nu_{i}$ is the number of valleys, $\sigma_i$ is the spin degeneracy of the band, $L^2$ is the area of the system, $\bm{k}_{F,i}$ is the Fermi wavevector, and $V_{\bm{q}} \chi_{\bm{q}}(\omega) = A_{\bm{q}}(\omega) + i \Sigma_{\bm{q}}(\omega)$. Here  $V_{\bm{q}}$ is the interaction potential and $\chi_{\bm{q}}(\omega)= \nu_{e} \chi_{e,\bm{q}}(\omega) + \nu_{h} \chi_{h,\bm{q}}(\omega)$ is sum of retarded electron and hole \textit{Lindhard} susceptibilities.
 
The finite thickness of Monolayer TMDCs makes it a quasi-2D system for which interactions between charge carriers are best described by the Keldysh potential \cite{Keldysh1978, MacDonald2015} which in Fourier space is of the form
\begin{equation}
\label{keldyshPotential}
V_{\bm{q}}=\frac{4\pi e^2}{(\epsilon_1 +\epsilon_2)q(1+r_0 q)}
\end{equation}
where $\epsilon_{1/2}$ is the dielectric constant of substrate above/below the quasi-2D layer,  and effective thickness $r_0=\epsilon d/ (\epsilon_1 +\epsilon_2)=4\pi \chi_{2D}/ (\epsilon_1 +\epsilon_2)=43.91 \mathrm{\AA}$ for free-standing MoS$_2$ \cite{Cudazzo2011,wang2014many} ($d$ is the thickness of the quasi-2D system, $\epsilon$ is the planar dielectric constant of the monolayer material, $\chi_{2D}$ is the 2D polarizability of the planer material). 
\begin{figure}[h]
 \centering
 \includegraphics[scale=0.45]{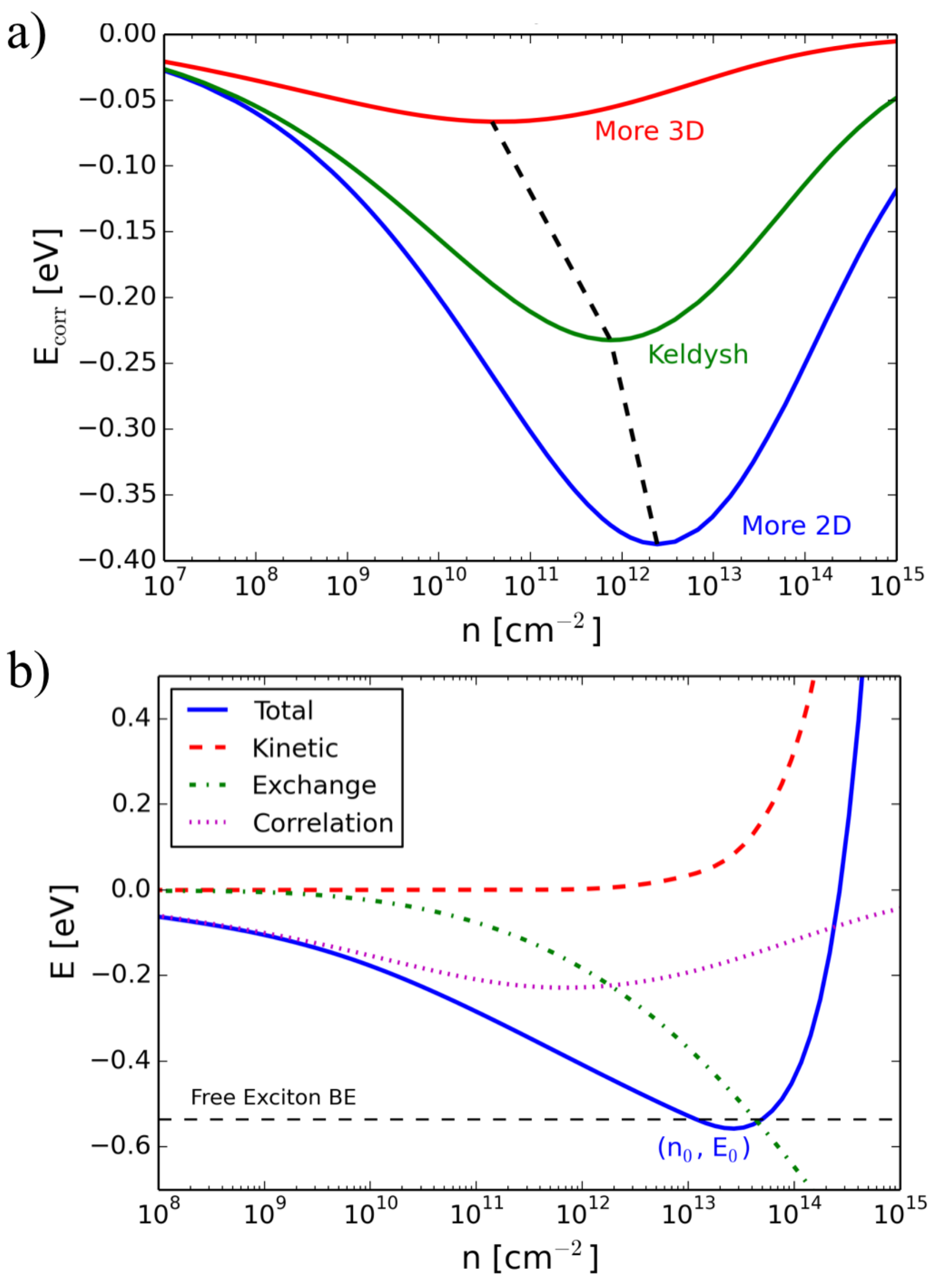}
 \caption{\label{zeroTEnergy} a) Correlation energy per electron-hole pair at T=0 K for different values of $r_0$. The Keldysh case corresponds to $r_0$= 43.91 $\mathrm{\AA}$ (MoS$_2$), `More 3D' case corresponds to $r_0$= 219.7 $\mathrm{\AA}$ ($V_q \sim 1/q^2$) and `More 2D' case corresponds to $r_0$= 21.96 $\mathrm{\AA}$ ($V_q \sim 1/q$). b) Ground state energy per electron-hole pair at T=0 K for MoS$_2$($r_0=$ 43.91 $\mathrm{\AA}$). The contributions from the kinetic, exchange, and correlation terms are marked.}
 \end{figure}
 
The Keldysh potential in equation (\ref{keldyshPotential}) has two significant effects. First, it provides the correct energy scale of few hundred meV for the exciton binding energy for monolayer TMDC's. Second, it accounts for the dimensional crossover of the interaction. The interaction behaves as three-dimensional ($\sim 1/q^2$) for small distances (large wavevectors) and two-dimensional ($\sim 1/q$) for large distances (small wavevectors) compared to the length scale provided by $r_0$ (see Fig.~\ref{bands}). The dimensional crossover of the interaction is clearly seen in the correlation energy calculation at T=0 K (see Fig.~\ref{zeroTEnergy}a) evaluated using the Keldysh potential. The density at which the interaction has dimensional crossover can be estimated as $r_0 k_\mathrm{F} \sim 1$ which implies $n \sim 1/2\pi r_0^2$. Fig.~\ref{zeroTEnergy}a shows that this dimensional crossover marked by the minima, happens at lower density for for higher $r_0$ and vice-versa. We also evaluate the contributions from the kinetic, exchange terms to get the variation of total ground state energy with density as shown in Fig.~\ref{zeroTEnergy}b following Eq.~\ref{ET0}. The lower ground state energy compared to the free exciton binding energy suggests EHL to be more stable. 

\section{Free Energy}
To map the phase diagram at finite temperature (i.e. effective temperature of the thermalized electrons and holes), we proceed to evaluate the thermodynamic potential $\Omega[\mu_e, \mu_h, T]$ at finite temperatures using the \textit{linked cluster expansion} method. The thermodynamic potential can be separated into three contributions, the non-interacting potential $\Omega_{0}$, the exchange potential $\Omega_\mathrm{X}$ and the correlation potential $\Omega_\mathrm{C}$ \cite{mahan2013many}. 
\begin{equation}
\label{linkedcluster}
\begin{split}
\Omega_{0} &= -\sum_{i=e,h}\frac{\nu_{i}\sigma_i}{\beta}  \sum_{\bm{k}} \log \left[1+\exp\left(-\beta(\varepsilon_{\bm{k},i}-\mu_{i} \right) \right]\\
\Omega_\mathrm{x} &= -\sum_{i=e,h}\frac{\nu_{i}\sigma_i}{2 L^2} \sum_{\bm{k},\bm{p}} V_{\bm{k}- \bm{p} }\, f_{i}(\bm{k}) f_{i}(\bm{p}) \\
\Omega_\mathrm{c} &= \frac{1}{2\beta} \sum_{i\omega_{n},\bm{q}} \bigg[ \log\big(1+ V_{\bm{q}} \chi_{\bm{q}}(i\omega_{n})\big) - V_{\bm{q}} \chi_{\bm{q}}(i\omega_{n})\bigg]
\end{split}
\end{equation}
where $\beta=1/k_\mathrm{B}T$ is the inverse thermal energy, $i\omega_{n}=2\pi n/\beta$ are the bosonic Matsubara frequencies, $\varepsilon_{\bm{k},i}=\hbar^2 k^2/2m_i$ is the energy dispersion (for MoS$_2$: $m_e=0.5$ $m_0$ and $m_h=0.59$ $m_0$ where $m_0$ is the free electron mass \cite{wang2014many}), $\mu_{i}$ is the chemical potential, $ f_{i}(\bm{k})$ is the Fermi-Dirac function, and $\chi_{\bm{q}}(i\omega_{n}) = \nu_{e} \chi_{e,\bm{q}}(i\omega_{n}) + \nu_{h} \chi_{h,\bm{q}}(i\omega_{n})$ is the sum of electron and hole susceptibilities. Each of the susceptibilities corresponding to electrons and holes is taken to be the \textit{Lindhard} susceptibility thus taking into account the contribution from ring diagrams. 

In the recent experimental observation of EHL in MoS$_2$ \cite{kenan}, the system is photoexcited with a CW laser beam. The non-equilibrium photoexcited  carriers attain dynamic equilibrium where the rate of carrier creation equals that of destruction. We note that by evaluating thermodynamic quantities, we are implicitly assuming ergodicity wherein the system has  attained dynamic equilibrium characterized by the steady state photoexcited density and the effective temperature of the thermalized carriers.

Since our photo-excited system has a severe constraint of having equal electron and hole densities $n_e=n_h=n$, it is appropriate for us to work in the canonical ensemble and consider the Helmholtz free energy $F[n,T]$ (also denoted as $A[n,T]$), of which the exchange-correlation contribution is related to the thermodynamic grand potential evaluated for chemical potential $\mu_{0,e/h}$ \cite{PerrotWardana1984}
 \begin{equation}
 \label{Helmholtz}
 F_\mathrm{xc}(n,T)\approx \Omega_\mathrm{xc}(\mu_{0,e},\mu_{0,h},T)
 \end{equation}
 where $\mu_{0,e/h}$ is the non-interacting chemical potential corresponding to the number density $n$. The chemical potential,
 which can show signatures of a phase transition,
  can be obtained from the Helmholtz free energy through an appropriate derivative,
 \begin{equation}
 \label{chemicalPotential}
\mu= \left( \frac{\partial F}{\partial N}\right)_{T,V} =\mu_{0,e}+\mu_{0,h}+ \left( \frac{\partial F_\mathrm{xc}}{\partial N}\right)_{T,V}.
 \end{equation}
\begin{figure}[h]
 \centering
 \includegraphics[scale=0.6]{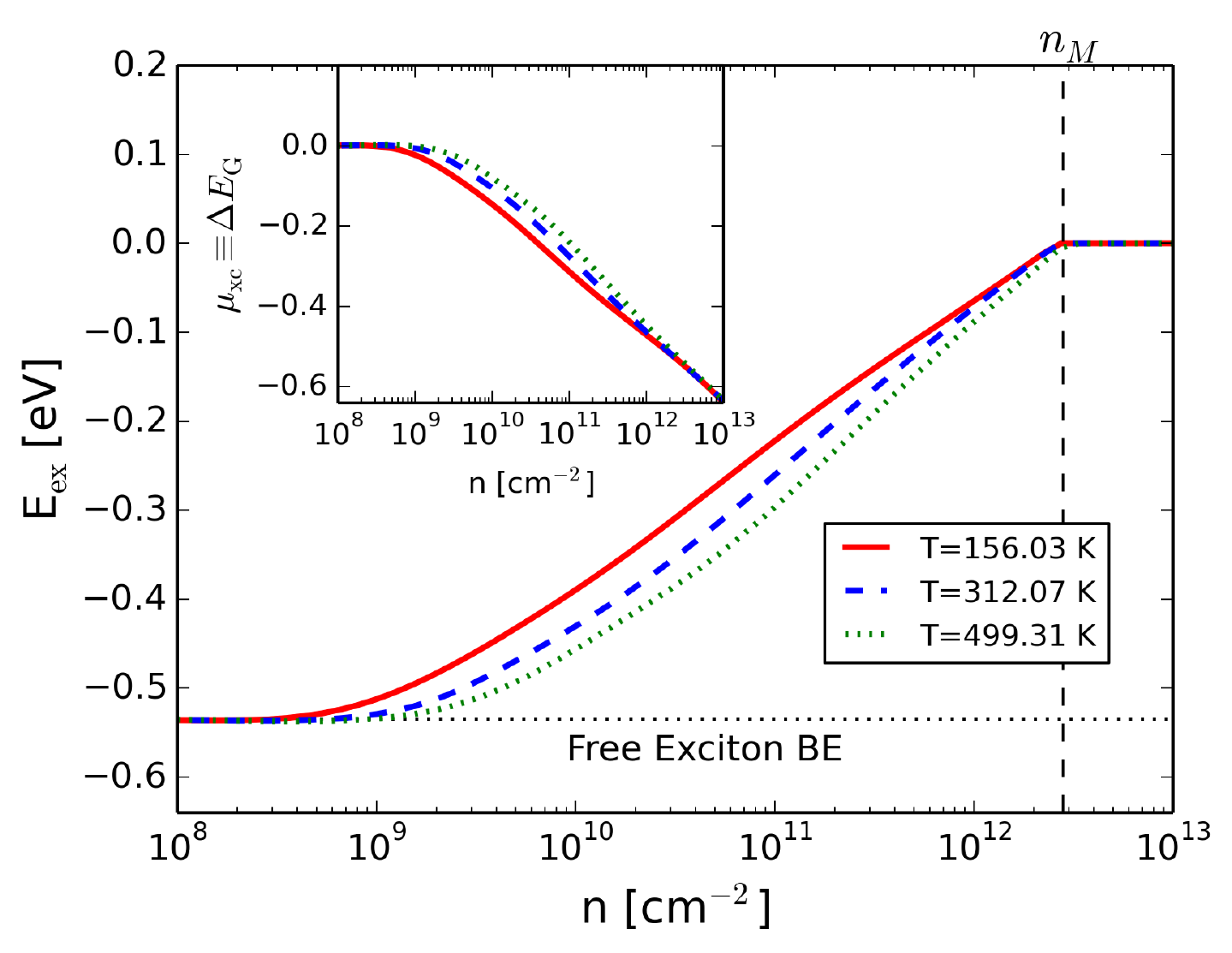}
 \caption{ \label{BGR} The variation of exciton binding energy with density for different temperatures. The inset shows band gap renormalization $\Delta E_{G} \equiv \mu_\mathrm{xc}$ as a function of density $n$ for different temperatures which is used to get density-dependent binding energy. The free exciton binding energy is denoted by the dashed line; the Mott transition occurs where these meet (at the Mott density  $n_M\approx 2.8 \times 10^{12}$ cm$^{-2}$).}
 \end{figure}
 \begin{figure*}[h]
 \centering
 \includegraphics[scale=0.56]{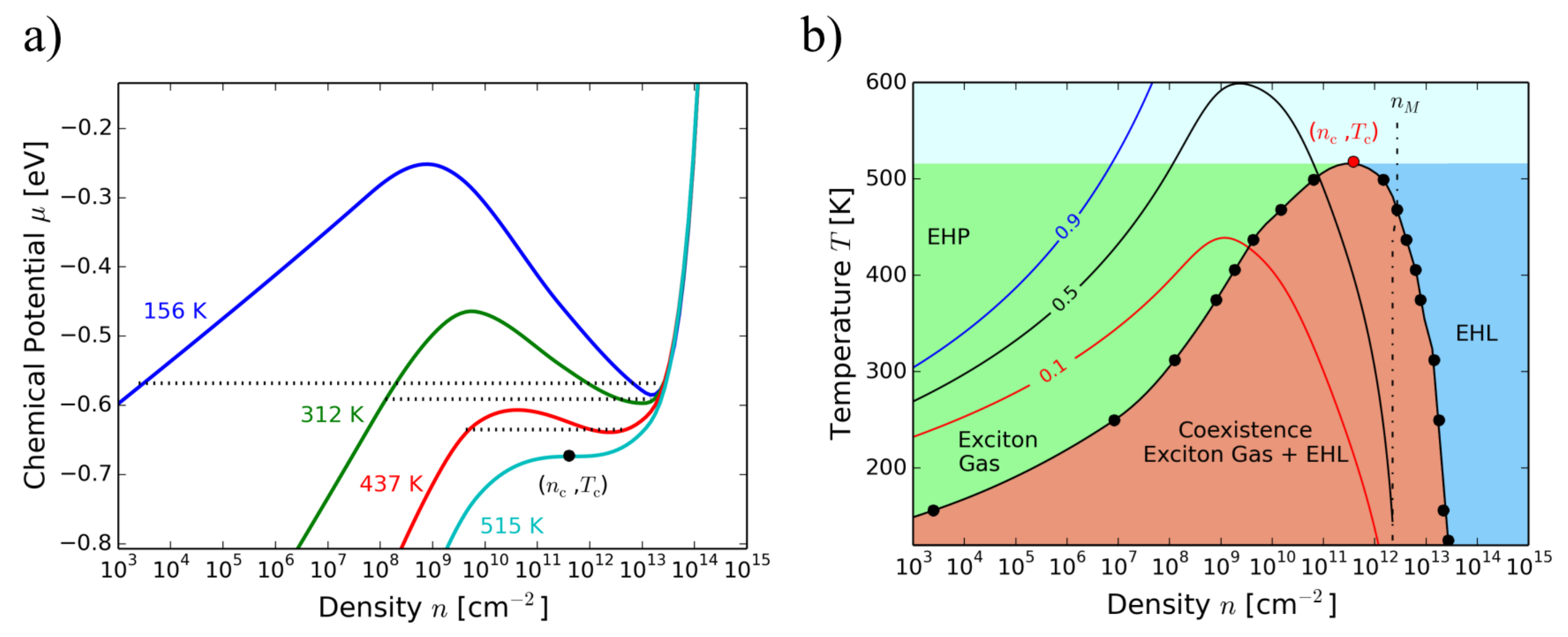}
 \caption{ \label{CombinedPD} a) Chemical potential $\mu$ as a function of number density of electron-hole pairs $n$ for different temperatures. The dashed lines indicate the Maxwell construction; the asymmetry in the area appears due to the density being plotted on a log scale. b) Phase Diagram in the temperature-density (T-n) plane indicating regions of gas, liquid and coexistence of
 electron-hole plasma (EHP), electron-hole liquid (EHL) and free excitons. The critical density n$_c$=3.8 $\times 10^{11}$ cm$^{-2}$ and the critical temperature T$_c$= 514.9 K. The figure also contains contour lines corresponding to ionization ratios $\alpha=\{0.1,0.5,0.9\}$ given by the Saha equation (equation (\ref{Saha})).}
 \end{figure*}
 
 \section{Band Gap Renormalization}
Photo-excited carriers have two major effects on the optical spectrum of a semiconductor. First, screening from the carriers weakens the interaction binding the electron-hole pairs causing a blue-shift of the exciton peak. Second, the self energy of the interacting system renormalizes the band gap and causes a red-shift of the exciton peak. Several studies have concluded that these energy shifts offset each other and the exciton peak location in the optical spectrum does not shift, however the continuum band edge shifts due to modification from band gap renormalization (BGR) \cite{BGRScreening_Haug1985}. Thus the variation in the exciton binding energy with density and temperature can be determined from the BGR. The exchange-correlation chemical potential may be used as an estimate of the band gap renormalization (BGR) \cite{RinkerBGR1992} (see inset in Fig.~\ref{BGR}) as a function of number density $n$ of electron-hole pairs at a few different temperatures. The BGR is used to determine the density and temperature dependent exciton binding energy (Fig.~\ref{BGR}). The Mott density where excitons merge into the continuum and dissociate into electron-hole plasma is found to be $n_\mathrm{M} \approx 2.8 \times 10^{12}$ cm$^{-2}$ independent of temperature. Our result is in close agreement with the exciton binding energy shift and band gap shift calculated by solving the semiconductor Bloch equations using an \textit{ab initio} band structure of MoS$_2$ \cite{Steinhoff_BEGapShiftSBE2014}.
 
\section{Phase Diagram} 
The variation of the chemical potential $\mu$ with density $n$ is shown in Fig.~\ref{CombinedPD}a for different temperatures. At high temperatures, the chemical potential increases monotonically with density; at low temperatures there exists a regime
in density where $\left(\partial\mu/\partial n\right)_{T,V} < 0$. This region is thermodynamically unstable and a Maxwell equal area construction (shown as horizontal dotted lines) is done to construct the region of coexistence. The discontinuity in the Maxwell constructed region classifies it as a \textit{first-order} phase transition for temperatures lower than the critical temperature T$_c$ . From this, we map out the phase diagram in temperature-density (T-n) space shown in Fig.~\ref{CombinedPD}b indicating regions of electron-hole plasma phase, liquid phase and the coexistence of the two. The coexistence region is characterized by a dense electron-hole plasma with droplets of constant high density electron-hole liquid. We find the critical density n$_c$=3.8 $\times 10^{11}$ cm$^{-2}$ and the critical temperature T$_c$= 514.9 K, which are indicated on the phase diagram.
 
Interactions lead to a first-order phase transition from a gas of electron-hole pairs to an electron-hole liquid. However, 
in actuality the gas phase is composed of exciton, biexcitons, and electon-hole plasma. 
Since our formalism does not incorporate all components in the gas phase, we use one of the most commonly used concepts in plasma physics,
the Saha ionization equation \cite{SahaEq_1920}, to describe the chemical equilibrium between electron-hole plasma and excitons $e+h \rightleftarrows X$
\begin{equation}
\label{Saha}
\dfrac{\alpha^2}{1-\alpha} = \dfrac{g_e g_h}{g_{ex}} \dfrac{1}{n \lambda_T^2} \exp\left( -\dfrac{\vert E_{ex}(n) \vert }{k_B T}\right),
\end{equation}
where $g$'s are the degeneracy factors ($g_e=4$ due to spin and valley degeneracy, $g_h=2$ due to valley degeneracy, and $g_{ex}=2$ due to valley degeneracy), $\alpha$ is the ionization ratio, $\lambda_T = h/\sqrt{2\pi m_r k_B T}$ is the thermal De Broglie wavelength of the
electron-hole pair, $m_r$ is the reduced mass of electron-hole pair. Fig.~\ref{CombinedPD}b displays ionization ratio lines corresponding to $\alpha=\{0.1,0.5,0.9\}$ based on equation(\ref{Saha}), incorporating the density-dependent exciton binding energy (ignoring the temperature dependence, using exciton binding energy for T=156 K).
As expected, at high temperatures and low density, the excitons dissociate into electron-hole plasma as the temperature becomes larger than the
exciton binding energy.  At low temperatures, the exciton gas is only partially dissociated into electron-hole plasma, but as the
density increases the plasma that is present coexists with the liquid phase (orange region).  As the density increases further, the
ionization ratio rapidly increases until the Mott transition is reached, beyond which only the liquid and plasma phases are found.

A widely used empirical relation exists relating the exciton binding energy $E_\mathrm{ex}$ and the critical temperature $T_\mathrm{c}$ ($k_\mathrm{B} T_\mathrm{c} \approx 0.1 E_\mathrm{ex}$) \cite{Keldysh1986} holds for our results. The family of monolayer TMDCs have exciton binding energies in the range of a few hundred meV \cite{Reichman_TMDCBE2013}. Thus it is expected that room temperature EHL state should be observable in the family of TMDCs. However, the presence of a substrate reduces the exciton binding energy\cite{MacDonald2015} and thus the critical temperature T$_\mathrm{c}$ should decrease as per the empirical scaling relation suggesting lower critical temperature in presence of a substrate compared to a free standing layer.

\section{Conclusions}
In conclusion, we map the phase diagram of monolayer MoS$_2$ showing EHL formation at and above room temperature. Such high critical temperature is caused by strong interactions described by the Keldysh potential given the quasi-2D nature of the system. The variation of the thermodynamically conjugate variables (chemical potential-density) shows discontinuity indicating a \textit{first-order} phase transition. We find the critical density to be n$_c$=3.8 $\times 10^{11}$ cm$^{-2}$ and the critical temperature to be T$_c$= 514.9 K for MoS$_2$. \\

\begin{acknowledgement}

The authors acknowledge the discussions and insights from A. Bataller, R. Younts and K. Gundogdu.

\end{acknowledgement}

%
%
%
%
\bibliography{EHL}

\end{document}